# Objective Information Theory: A Sextuple Model and 9 Kinds of Metrics

Xu Jianfeng[①]* Tang Jun[②] Ma Xuefeng[①] Xu Bin[①] Shen Yanli[①] Qiao Yongjie[①]

① China Academy of Electronics and Information Technology
② Tsinghua University
*E-mail: xjf.av.sss@tom.com

**Abstract**

In the contemporary era, the importance of information is undisputed, but there has never been a common understanding of information, nor a unanimous conclusion to the researches on information metrics. Based on the previous studies, this paper analyzes the important achievements in the researches of the properties and metrics of information as well as their main insufficiencies, and discusses the essence and connotation, the mathematical expressions and other basic problems related to information. On the basis of the understanding of the objectivity of information, it proposes the definitions and a Sextuple model of objective information; discusses the basic properties of information, and brings forward the definitions and mathematical expressions of nine kinds of metrics of information, i.e., extensity, detailedness, sustainability, richness, containability, delay, distribution, validity and matchability. Through these, this paper establishes a basic theory frame of Objective Information Theory to support the analysis and research on information and information system systematically and comprehensively.

**Keywords**    Information    Objective Information Theory    Sextuple Model    Metrics

## 1    Introduction

With the rapid development of computer and communication technology, human society has entered into the information age, and the ubiquitous information systems have brought tremendous changes to people's production and life style. However, there has never been a common understanding of information, nor a unanimous conclusion to the researches on information metrics. In essence, this is not due to people's lack of interest or effort, but caused by the absence of an overarching and pervasive consensus about the object, goal and method of information study among various researchers. Firstly, about the object of information study, as there are both objective and subjective elements in what people commonly understand as information [1], many studies do not distinguish between the objectively-existent information and human subjective consciousness, which make the issue extremely complicated; secondly, about the goal of information study, a lot of studies are satisfied with an academic exploration and deduction, but have overlooked, to a great extent, the various information systems in the objective world that are alternating or will alternate people's production and life style profoundly; thirdly, about the method of information study, many studies only inherit Shannon's traditional theoretical approach of probability statistics with regard to communication systems, but have neglected the feasible approach of establishing information science theoretical basis systematically and comprehensively with the more extensive mathematical theory.

Given that, this paper analyzes and learns from the profound research achievements made by predecessors, and brings forward the basic frame of Objective Information Theory, including:

Firstly, the essence of information is explicitly confined within the objective realm, which is in accordance with both the position of information as one of the three fundamental elements of the objective world, and the operating requirements that information systems can and only can handle objectively-existent information.

Secondly, the mathematical expressions and Sextuple model of Information are brought forward, providing future studies with a unified, clear and applicable mathematical theory.

Thirdly, the basic properties of information, such as objectivity, reducibility, transitivity, relevance and combinability, are elaborated and verified based on the Sextuple Model and the mathematical theory of mapping, which further proves the reasonableness of the definitions of information and its Sextuple model.

Fourthly, the systematic and comprehensive metric system of Information with nine kinds of metrics is established, with a description of the mathematical expressions and basic properties of each metric on basis of fundamental models. It provides the theoretical and methodological basis for quantitative analysis of information working and information system application.

## 2  Several Key Issues

The establishment of a basic theoretical system of information must be based on the essence and connotation of information, and should start with mathematical models, so as to form an integrated metrics system and the accordant expressions. A large amount of constructive researches have been done before, which deserve to be well analyzed and summarized.

### 2.1  About the Essence and Connotation of Information

Information as a scientific term was first found in *Transmission of Information* published by Harley in 1928 [2]. It is generally acknowledged that *A Mathematical Theory of Communication* [3] published by Shannon in 1948 laid the foundation for modern information theory, in which Shannon described information as the random codes that needed to be transmitted during communication process and obeyed a certain statistical distribution. Thence the essence of communication technology was recognized and grasped for the first time in history，having promoted the flourishing development of communication system and brought great convenience to exchanging ideas. Along with the development of information technology and information systems, the existence and application of information goes far beyond communication system and the understanding of information grows so rich and profound that "there have appeared over one hundred popular theories on the definition of information" [4], which generally make analogies with the concepts of message, signal, data, intelligence, knowledge and so on. Each concept has its own focus and is inevitably partial, so "it is far from reaching a consensus on the essence of 'information' now in academia" [5].

Among numerous researchers, Zhong Yixin has done the most abundant research on the concept of information [4]. The *Comprehensive Information Theory* built by Zhong defines information at the ontology level as "the self-expression and self-display of objects' motion states and ways of state changing". Meanwhile, it holds that only when its formation is apperceived, its meaning is understood, and its value is determined, can information at epistemology level be really comprehended. Thereinto, the formation factor contained in information is termed *syntactical information*, the meaning factor is termed *semantic information* and the utility factor is termed *pragmatic information*. The epistemology-level information containing exterior formation, interior meaning and utility value of objects' motion states and ways of state changing is termed *comprehensive information*. The *Comprehensive Information Theory*'s semantic and pragmatic study of information has accomplished the mission that information scientists of the West have not finished [6], but its completeness has also exposed itself to various obstacles. Firstly, it does not put emphasis on the *information at ontology level* which is objectively existent and has relatively simple mechanism. Secondly, its major concern of the *information at epistemology level* is closely related to human subjective thought, which has very complicated principles and the research of which has to bridge the *Semantic Gap* that has not been well solved till now. Thirdly, it attempts to unify the Recognition Theory, Communication Theory, Control Theory, Decision Theory, Optimization Theory and Intelligence Theory into one integrated theoretical system and there are still many intricate issues to be clarified.

Burgin establishes the general theory of information on the basis of ontological and axiological principles, defining information as the capacity to cause changes [7]. Meanwhile, he introduces the concept of an *infological system $IF(R)$* of the system R and points out that information is relative, i.e., information for a system $R$ is only the capacity to cause changes in the system $IF(R)$. The idea of building a complete information theory with the general theory of information based on axiomatic characterizations is recommendable, but the major problems with this approach lie in: firstly the definition of information as the capacity to cause changes brings energy into the realm of information, which is quite different to people's common knowledge; secondly, the system $R$ and its infological system $IF(R)$ include both objective and subjective systems, leaving it still difficult to distinguish between information and human subjective perception.

Fleissner and Hofkirchner further bring forward the concept of Unified Information Theory [8], which, based on an analysis of the evolutionary process of information from syntactical to semantic and to pragmatic states, describes information as the product of changes in the system's structure, state or output. Unified Information Theory tries to use a consistent concept to unify all of the different connotations of information, but it is similar to Zhong's Comprehensive Information Theory in terms of basic analysis while far less profound and systematic than the latter. Therefore, even the founders of Unified Information Theory have doubt about its feasibility, and points out the major obstacles to building a unified theory [9].

2.2 About the Mathematical Expressions of Information

In reference [3], although Shannon does not directly give the mathematical definition of information, he does reveal that *information is a set of codes with occurrence probabilities denoted by* $p_1, p_2, \cdots, p_n$ respectively and the entropy formula

$$H(X) = H(p_1, p_2, \cdots, p_n) = -k \sum_{i=1}^{n} p_i \log p_i \tag{2.1}$$

is the metric to reflect the indeterminacy of information, and then, the basic principles of communication system design are established.

Zhang Jiguo illustrates the concept of information point [10], and gives its mathematical expression

$$R = (N, C, v). \tag{2.2}$$

Here, $N$ denotes information, $c$ denotes the name of information characteristics and $v$ denotes the metric of $N$ on $c$. Based on such sequential triple model, achievements on information field, information group and information potential have been acquired.

Zhang Weiming brings forward a mathematical model of information systems [11]

$$S = \langle O, A, V, \rho \rangle. \tag{2.3}$$

Here, $O$ is the finite set of objects, $A$ is the finite set of object attributes, $V$ is the union of the definition domain $V_a$ of each attribute, i.e. $V = \bigcup_{a \in A} V_a$, and $\rho$ is the attribute function of objects. Based on this model, achievements have been made in the research of the most complete information system, the simplest information system, the ideal information systems, and the indeterminate information systems and the connection among information systems.

These researches all try to apply mathematical expressions or modeling to the study of information. Thereinto, Shannon put forward the mathematical expressions of information, which are mainly concerned with statistical information in communication systems. Although his achievement is distinguished and widely applied in other areas, it is quite different from the mathematical connotation of general information. The mathematical expression of information point is most similar to the concept of general information, but its component parameters are neither related to the practical definition of information, nor able to play a crucial role in further researches. Though the mathematical model of information systems does not directly aim at information, it is very close to the component factors of information and its parameters are well applied in follow-up studies. This approach deserves to be well utilized.

The authors of this paper consider the Triple model established by Burgin in the general theory of information to be the most important achievement ever made in the researches on the mathematical expressions of information [12]

$$\langle C, I, R \rangle. \tag{2.4}$$

Here, $C$ denotes carrier, $I$ denotes information and $R$ denotes information system. Applying the category theory method on the basis of this model, through a series of operators among $C$, $I$, $R$ and infological system $IF(R)$ [12], Burgin studies the relationship among $C$, $I$ and $R$ during information sending, transmitting, receiving and saving. This method regards information, its carrier and the system it belongs to as a complete system, which greatly facilitates the study on the essence and operation of information with various mathematical methods and should be applied to all researches on information as a basic method. However, this Triple model is not strictly correspondent with the definition of information given in the general theory of information. In this model, information $I$, which should be the research object, is not further decomposed; information system $R$, which should be independent from information, becomes closely related with $I$; carrier $C$, which should be a component of information, is not included in $I$; and infological system $IF(R)$, which should be an important factor in the general information theory's definition of information, is not reflected in the Triple model. Therefore, the Triple model of the general information theory fails to fully represent the basic structure of all elements contained in information and is actually not fully related with the information operators research based on category theory.

2.3 About the Metrics of Information

Hartley is the first to put forward the method to measure the quantity of information from information source in transmission [2]. The main idea is that if we select $n$ symbols randomly from $S$ different symbols to compose a character, and the occurrence probabilities of all symbols are equal, then $S^n$ different characters will be acquired and the quantity of information contained in them is as follows

$$H = nlogS. \qquad (2.5)$$

Though the pure formalized method to measure the quantity of information set up by Hartley is too simple, it has imposed immediate influences on Shannon. Entropy formula (2.1) is obviously the evolution and expansion of (2.4). As the metric of information, it marked the birth of information theory. Thereafter, Wiener [13]、Ashby [14]、De Luca and Termini [15] have all tried to describe information metrics with their own Entropy formulas. Large amount of following studies have successively introduced the concepts of the expansion of Hartley entropy and Shannon entropy [16], relative entropy [17], cumulative residual entropy [18-21], joint entropy [22,23], conditional entropy[24-26], mutual information [27-32], cross entropy [33-38], fuzzy entropy [15,39], maximum entropy principle [40,41] and minimum cross-entropy principle [42,43], and a series of achievements have been made in these aspects. Zhong makes use of general information functions to unify the methods of describing information metrics with Entropy formulas [4]. It is necessary to emphasize that these researches on entropy mainly applies the mathematical method based on probability statistics.

In 1945, Rao put forward the concept of statistical manifold [44] and introduced a Riemannian metric in a model of probability distribution with the help of Fisher information matrix. He applied geodesic distance, also called Fisher information distance, to measuring the difference between probability distribution functions, which formed a contrast with Shannon's information entropy and became another means of reflecting information metrics. The effective combination of probability statistics and geometric methods led to the beginning of information geometry theory [45]. Thereafter, Efron studied the curvature of statistical manifold [46], Chentsov introduced a family of affine connections [47], and Amari put forward the concept of dual affine connections [48-50] which became a weaker distance function by introducing potential function to define the concept of divergence. In recent years, information geometry has been widely applied in various domains, including information theory, system theory, neural network, statistical inference, communication coding, physics and medical imaging [51-62]. As a useful tool for studying information metrics, information geometry is considered *the second generation of Information Theory* [63]. However, the research on this theory and its applications is still in the beginning stage due to short time of development. In addition, the geometric space in mathematics is just a form of measure space or topological space and the various metrics of information are not necessarily able to be fully reflected by their projection on the geometric space, which necessitates the introduction of more extensive mathematical tool.

Burgin's general information theory also involves the issue of information metrics [64]. It defines three kinds of structural information metrics—internal, intermediate and external—respectively according to objects and degree of changes, and three kinds of constructive information metrics—abstract, realistic and experimental—respectively according to different means of acquisition. These definitions have no specific mathematical expressions and therefore are only meaningful in terms of conceptual study.

Obviously, the function of information is not limited within the communication domain, and the metrics of information are by no means limited to the quantity of information required for eliminating indeterminacy or causing changes. Therefore, a lot of beneficial exploration of other kinds of information metrics has been made.

The metric similar to quantity of information is information richness. Information Richness Theory, also known as Media Richness Theory [65] and first put forward by Daft and Lengel in 1986, expands the focus of concern from communication and one-way information transmission to media and two-way information interaction respectively. The research approach of Information Richness Theory is entirely different from Shannon's information theory, but the way to measure the value and capability of media with information richness has greatly enlightened the process of building up and operating the metrics system of information.

The detailedness of information is also an important metric. Liang defines information entropy, rough entropy and knowledge granularity [66] of complete and incomplete information systems, and gives the axiomatic definition of information granularity, which is an average metric of the different levels of information detailedness. And then, Teng continues to study the relationship between entropy and information granularity [67], and points out that information granularity in information systems reflects the aggregation degree of knowledge to classifications of universe of discourse, i.e., the rougher the classifications are, the larger the information granularity is; and conversely the smaller the information granularity is. His research further clarifies the meaning of the metric of information detailedness.

The value of information lies in its validity to a large extent, so the metric of validity has been highly emphasized, especially in military domain. Gaziano and McGrath have done some researches on the concept of creditability metric [68]; Alberts [69] and RAND Corporation [70] have also studied the metric definition and estimation method of information correctness and creditability according to the requirements of military application. Then Niven investigates the creditability of information in media and news [71]. In fact, the creditability and correctness here both refer to the difference between information and the real state of objects reflected by it, so validity should be the most important metric of information.

Moreover, timeliness, universality and satisfiability to users' demand are all indispensible metrics of information. Alberts [69] and RAND Corporation [70], etc. have all done some researches on these metrics, which provides helpful references for constructing the metric system of information.

The main deficiency of the previous researches is lack of systematicness. Firstly, they did not take objectivity of information as the basis of their research on information metrics. Secondly, most of them focused on only one metric and lacked overall consideration. Thirdly, most of them did not give the corresponding mathematical expressions of the definitions of information metrics, making further application very difficult. Fourthly, the majority of metrics proposed in these researches can only be applied in specific fields and scenarios and therefore, lack universality.

3  The Definitions and Properties of Information

It may be a very difficult philosophical issue to define information. This paper is not intended to unify all the different opinions, but just brings forward the definitions, mathematical expressions and basic properties of objective information from a perception approved by the authors and in accordance with the operational requirements of information systems.

3.1  The Definition of Information

It is commonly acknowledged that the world consists of the subjective world and the objective world. The subjective world is a world of human consciousness and notions, which is the summation of people's mental and psychological activities during perceiving and understanding the whole world. The objective world is a perceptible and material world which is the summation of all matters and their motions except for human beings' consciousness activities.

N. Wiener, the founder of Control Theory, is the first in history to raise the position of information to object of study. He points out that *information is information, neither matter nor energy* [13]. This viewpoint is called the Matter-Energy-Information Trichotomy about the world (here the "world" actually refers to the objective world in the authors' understanding), which holds that Matter, Energy and Information are the three element factors constituting the objective world [72]. Thereinto, matter is an objective reality independent of human subjective consciousness; energy is the motion ability of matter; and information is the objective reflection of *objects and their motion states* in the subjective and the objective world through the media of matter or energy. In short, matter is the being of origin, energy is the being of motion ability and information is the being of relations among objects. All kinds of physical information systems can and only can acquire, transmit, process and apply the information of objective existence. So the Trichotomy meets the objectivity requirements for the objects of information system operation. All these viewpoints are completely based on the objective reality of information and are thus called Objective Information Theory. So the following definitions of Information can be acquired.

**Definition 1** Information is the objective reflection of *objects and their motion states* in both the objective and the subjective world.

According to this definition, the following specific concepts can be deduced.

**Definition 2** (Information Ontology) Information reflects *objects and their motion states* in both the objective and the subjective world, and the set of objects here is termed Information Ontology, Ontology for short.

**Definition 3** (Information State Occurrence Time) Information reflects *objects and their motion states* in a specific time set. The specific time set here is termed Information State Occurrence Time, State Occurrence Time for short.

**Definition 4** (Information State Set) Information reflects *objects and their motion states*. The set of *motion states* here is termed Information State Set, State Set for short.

**Definition 5** (Information Carrier) Information can not reflect anything without certain media set in the objective world. The set of media here is termed Information Carrier, Carrier for short.

**Definition 6** (Information Reflection Time) Information exists in Carrier in a set of specific time. The time set here is termed Information reflection Time, Reflection Time for short.

**Definition 7** (Information Reflection Set) Information itself exists in the time-space of Carrier as a series of states. The series of states is termed Information Reflection Set, Reflection Set for short.

3.2 The Mathematical Expressions of Information

Let $O, S$ and $T$ be the sets of objective world, subjective world and time respectively. The elements in $O, S$ and $T$ can be divided properly according to specific requirements of domain of discourse. So we can get the mathematical definition of Information:

**Definition 1'** Assuming that Ontology $o \in 2^{O \cup S}$, State Occurrence Time $T_h \in 2^T$, State Set of $o$ on $T_h$ as denoted by $f(o, T_h)$, Carrier $c \in 2^O$, Reflection Time $T_m \in 2^T$ and Reflection Set of $c$ on $T_m$ as denoted by $g(c, T_m)$ are all nonvoid, then Information $I$ is the surjective mapping from $f(o, T_h)$ to $g(c, T_m)$. Namely,

$$I: f(o, T_h) \to g(c, T_m), \tag{3.1}$$

or
$$I(f(o, T_h)) = g(c, T_m). \tag{3.2}$$

All the sets of information $I$ are called an information space, denoted by $\mathfrak{T}$, which is one of the three element factors of the objective world.

Therefore, not only the mapping $I$, but also Ontology $o$, State Occurrence Time $T_h$, State Set $f(o, T_h)$, Carrier $c$, Reflection time $T_m$ and Reflection Set $g(c, T_m)$ are all very important for information.

**Definition 8** $\langle o, T_h, f, c, T_m, g \rangle$ is the Sextuple model of Information $I$, denoted by $I = \langle o, T_h, f, c, T_m, g \rangle$.

3.3 The Basic Properties of Information

Based on the Sextuple model and the mathematical theory of mapping, we could further discuss the basic properties of information.

(1) Objectivity

According to Definition 1', Information $I = \langle o, T_h, f, c, T_m, g \rangle$ is the surjective mapping from $f(o, T_h)$ to $g(c, T_m)$, so Information $I$ can only be reflected through $g(c, T_m)$ which is the State Set of Carrier $c$ in the objective world on Reflection Time $T_m$. Therefore, Information $I$ can only be reflected through the objective world. This is the constraint condition for information perception.

(2) Reducibility

Information $I = \langle o, T_h, f, c, T_m, g \rangle$ is the surjective mapping from $f(o, T_h)$ to $g(c, T_m)$, so there naturally exists the inverse mapping of $I$:

$$I^{-1}: g(c, T_m) \to f(o, T_h), \tag{3.3}$$

or

$$I^{-1}(g(c, T_m)) = f(o, T_h). \tag{3.4}$$

That is to say, State Set of $o$ on $T_h$ as denoted by $f(o, T_h)$ could be reduced from $g(c, T_m)$ and $I^{-1}$. This is the fundamental premise for information application.

(3) Transitivity

Information $I = \langle o, T_h, f, c, T_m, g \rangle$ is the surjective mapping from $f(o, T_h)$ to $g(c, T_m)$. It is certainly possible that there exist Carrier set $c'$ in the objective world, time set $T'_m$ and set of all states of $c'$ on $T'_m$ as denoted by $g'(c', T'_m)$, which form a surjective mapping from $g(c, T_m)$ to $g'(c', T'_m)$. Here, according to the definitions, the mapping

$$I': g(c, T_m) \to g'(c', T'_m) \tag{3.5}$$

is also information, and

$$I'(g(c, T_m)) = I'(I(f(o, T_h))). \tag{3.6}$$

is actually the compound mapping from $f(o, T_h)$ to $g'(c', T'_m)$ through $g(c, T_m)$.

Therefore, compound mapping $I'(I(f(o, T_h)))$ accomplishes the transmission of information from $o$, $T_h$

and $f(o, T_h)$ to $c$, $T_m$ and $g(c, T_m)$, and then to $c'$, $T_m'$ and $g'(c', T_m')$. This is the basic mode of information transmission.

(4) Relevance

For Information $I = \langle o, T_h, f, c, T_m, g \rangle$, $o$ and $c$, $T_h$ and $T_m$, and particularly $f(o, T_h)$ and $g(c, T_m)$ all come in pairs. As the surjective mapping from $f(o, T_h)$ to $g(c, T_m)$, Information $I$ establishes a certain relationship between the states of $o$ and $c$. Due to the property of relevance, information could relate more objects together. This is where the significance of information lies.

(5) Combinability

In Information $I = \langle o, T_h, f, c, T_m, g \rangle$, $o, T_h, f, c, T_m, g$ are sets with different roles, which can be decomposed or combined to several new sets, so information possesses the property of combinability.

**Definition 9** (Sub-information) For Information $I' = \langle o', T_h', f', c', T_m', g' \rangle$ and $I = \langle o, T_h, f, c, T_m, g \rangle$, if
$$o' \subseteq o, T_h' \subseteq T_h, f' \subseteq f, c' \subseteq c, T_m' \subseteq T_m, g' \subseteq g \tag{3.7}$$
and
$$I'(f'(o', T_h')) = I(f'(o', T_h')), \tag{3.8}$$
then $I'$ is called sub-information of $I$, denoted by
$$I' \subseteq I. \tag{3.9}$$
Moreover, if at least one of the following expressions
$$o' \subset o, T_h' \subset T_h, f' \subset f, c' \subset c, T_m' \subset T_m, g' \subset g \tag{3.10}$$
holds, then $I'$ is called proper sub-information of $I$.

**Definition 10** (Combined Information) For Information $I = \langle o, T_h, f, c, T_m, g \rangle$ and proper sub-information $I' = \langle o', T_h', f', c', T_m', g' \rangle$ and $I'' = \langle o'', T_h'', f'', c'', T_m'', g'' \rangle$, if
$$o = o' \cup o'', T_h = T_h' \cup T_h'', f = f' \cup f'', c = c' \cup c'', T_m = T_m' \cup T_m'', g = g' \cup g'', \tag{3.11}$$
and for any $o_\lambda \in o, T_{h\lambda} \in T_h, f_\lambda \in f, c_\lambda \in c, T_{m\lambda} \in T_m$ the following expression
$$I(f(o_\lambda, T_{h\lambda})) = I'(f'(o_\lambda, T_{h\lambda})) \text{ or } I''(f''(o_\lambda, T_{h\lambda})) \tag{3.12}$$
holds, then $I$ is the combined Information of $I'$ and $I''$, denoted by
$$I = I' \cup I''. \tag{3.13}$$

4  The Metrics of Information

The establishment of the metrics system of Information must be based on the following principles:

(1) The principle of Traceability—to determine the specific definitions and mathematical expressions of all kinds of metrics according to the definition model of Information;

(2) The principle of Completeness—to form a complete metric system which is closely related to its value according to the practical connotation of Information;

(3) The principle of Universality—to form metric definitions based on various categories of Information which is universally applicable to information acquisition, information transmission, information processing, information application and any collection of these information systems rather limited to a specific realm;

(4) The principle of Practicability—to form a practical and workable metric system which could guide the analysis and research of information system according to the application requirements of information.

Therefore, the information metric system based on Sextuple model is comprised of the following nine kinds of specific definitions and related basic propositions that could be easily derived from properties such as set measure, potential and distance.

4.1 The Metric of Information Extensity

Since Information is the objective reflection of Ontology, its value is directly related to the coverage of Ontology. Generally speaking, when other conditions are all the same, the wider the Ontology coverage is, the higher the Information value is; and conversely the lower the value is. Therefore, Information could be measured by Extensity or Scope, the mathematical expression for which is:

**Definition 12** (Information Scope) Assuming that $O$ and $S$ are the sets of objective and subjective world respectively, $(O \cup S, 2^{O \cup S}, \mu)$ is a measure space, and $\mu$ is some measure on the measurable set $2^{O \cup S}$, then the scope of information $I = \langle o, T_h, f, c, T_m, g \rangle$ with respect to measure $\mu$, denoted by $scope_\mu(I)$, is the measure of $o$ on the measurable set $2^{O \cup S}$, i.e.

$$scope_\mu(I) = \mu(o). \tag{4.1}$$

**Proposition 1** (Total scope is wider than partial scope) For Information $I = \langle o, T_h, f, c, T_m, g \rangle$ and $I' = \langle o', T_h', f', c', T_m', g' \rangle$, if $I'$ is sub-information of $I$, then

$$scope_\mu(I') \leq scope_\mu(I). \tag{4.2}$$

4.2 The Metric of Information Detailedness

The value of Information is also directly related to the detailedness of the Ontology it reflects. Detailedness indicates the degree of coarseness of the particles that the Ontology can be decomposed into. Generally speaking, when other conditions are all the same, the finer the particles are, the higher the Information value is; and conversely the lower the value is. Therefore, Information could be measured by Detailedness or Granularity, the mathematical expression for which is:

**Definition 13** （Atomic Information）For Information $I = \langle o, T_h, f, c, T_m, g \rangle$ and $I' = \langle o', T_h', f', c', T_m', g' \rangle$, if $I'$ is proper sub-information of $I$, and there is no other proper sub-information $I'' = \langle o'', T_h'', f'', c'', T_m'', g'' \rangle$ of $I$, such that

$$I'' \subset I', \tag{4.3}$$

then $I'$ is called atomic information of $I$.

**Definition 14** (Information Granularity) Assuming that $O$ and $S$ are the sets of objective and subjective world respectively, $(O \cup S, 2^{O \cup S}, \mu)$ is a measure space, $\mu$ is a some measure on the measurable set $2^{O \cup S}$, and the set of all the atomic information of Information $I = \langle o, T_h, f, c, T_m, g \rangle$ is denoted by $A = \{I_\lambda = \langle o_\lambda, T_{h\lambda}, f_\lambda, c_\lambda, T_{m\lambda}, g_\lambda \rangle\}_{\lambda \in \Lambda}$, where $\Lambda$ is an index set, then the granularity of information $I$ with respect to measure $\mu$, denoted by $granularity_\mu(I)$, is the maximum value of the measures of all Ontologies in $A$, i.e.

$$granularity_\mu(I) = max_{\lambda \in \Lambda}\{\mu(o_\lambda)\}. \tag{4.4}$$

**Proposition 2** (Total granularity is coarser than partial granularity) For Information $I = \langle o, T_h, f, c, T_m, g \rangle$ and $I' = \langle o', T_h', f', c', T_m', g' \rangle$, if all the atomic information of $I'$ are also the atomic information of $I$, then

$$granularity_\mu(I') \leq granularity_\mu(I). \tag{4.5}$$

4.3 The Metric of Information Sustainability

The value of Information is also related to the sustainability of the Ontology it reflects. Sustainability indicates the density and span of Occurrence Time. Generally speaking, when other conditions are all the same, the higher the density is and the longer the span is, the higher the Information value is; and conversely the lower the value is. Therefore, Information could be measured by Sustainability, the mathematical expression for which is:

**Definition 15** (Information Sustainability) Assuming that $T$ is a time set, $(T, 2^T, \tau)$ is a measure space, and $\tau$ is some measure on the measurable set $2^T$, then the sustainability of Information $I = \langle o, T_h, f, c, T_m, g \rangle$ with respect to measure $\tau$, denoted by $sustainability_\tau(I)$, is the measure of $T_h$ on the measurable set $2^T$, i.e.

$$sustainability_\tau(I) = \tau(T_h). \tag{4.6}$$

**Proposition 3** (Total sustainability is greater than partial sustainability) For Information $I = \langle o, T_h, f, c, T_m, g \rangle$ and $I' = \langle o', T_h', f', c', T_m', g' \rangle$, if $I'$ is sub-information of $I$, then

$$sustainability_\tau(I') \leq sustainability_\tau(I). \tag{4.7}$$

4.4 The Metric of Information Richness

The value of Information is also related to the richness of its reflection of the Ontology State, which is actually embodied in the amount of the content contained in the Reflection Function. Generally speaking, when other conditions are all the same, the richer the content is, the higher the Information value is; and conversely the lower the value is. Therefore, Information could be measured by Richness, the mathematical expression for which is:

**Definition 16** (Information Richness) Assuming that $\mathcal{F}$ is the set of all State Functions of Ontology $o$ on $T_h$, $(\mathcal{F}, 2^\mathcal{F}, \rho)$ is a measure space, and $\rho$ is some measure on the measurable set $2^\mathcal{F}$, then the richness of Information $I = \langle o, T_h, f, c, T_m, g \rangle$ with respect to measure $\rho$, denoted by $richness_\rho(I)$, is the measure of $f(o, T_h)$ on the measurable set $2^{\mathcal{F}}$, denoted by $\rho(f(o, T_h))$, i.e.

$$richness_\rho(I) = \rho(f(o, T_h)). \tag{4.8}$$

**Proposition 4** (Total richness is more than partial richness) For Information $I = \langle o, T_h, f, c, T_m, g \rangle$ and $I' = \langle o', T_h', f', c', T_m', g' \rangle$, if $I'$ is sub-information of $I$, then

$$richness_\rho(I') \leq richness_\rho(I). \tag{4.9}$$

4.5 The Metric of Information Containability

The value of Information is also closely related to its inherent Containability, which is reflected by the required volume for the Carrier. Generally speaking, when other conditions are all the same, the smaller the required volume is, the higher the Information value is; and conversely the lower the value is. Therefore, Information could be measured by Containability or Volume, the mathematical expression for which is:

**Definition 17** (Information Volume) Assuming that $O$ is the set of objective world, $(O, 2^O, \sigma)$ is a measure space and $\sigma$ is some measure of the measurable set $2^O$, then the volume of Information $I = \langle o, T_h, f, c, T_m, g \rangle$ with respect to measure $\sigma$, denoted by $volume_\sigma(I)$, is the measure of $o$ on the measurable set $2^O$, denoted by $\sigma(c)$, i.e.

$$volume_\sigma(I) = \sigma(c). \tag{4.10}$$

**Proposition 5** (Total volume is larger than partial volume) For Information $I = \langle o, T_h, f, c, T_m, g \rangle$ and $I' = \langle o', T_h', f', c', T_m', g' \rangle$, if $I'$ is sub-information of $I$, then

$$volume_\sigma(I') \leq volume_\sigma(I). \tag{4.11}$$

It is not hard to prove that Shannon's Information entropy is actually the indicator of information volume required for delivering discrete messages in communication systems.

4.6 The Metric of Information Delay

The value of Information is certainly related to its timeliness, which is embodied in the delay between Reflection Time and State Occurrence Time. Generally speaking, when other conditions are all the same, the shorter the delay is, the higher the Information value is; and conversely the lower the value is. Therefore, Information could be measured by Delay, the mathematical expression for which is

**Definition 18** (Information Delay) The set of all the atomic information of information $I = \langle o, T_h, f, c, T_m, g \rangle$ is denoted by $A = \{I_\lambda = \langle o_\lambda, T_{h\lambda}, f_\lambda, c_\lambda, T_{m\lambda}, g_\lambda \rangle\}_{\lambda \in \Lambda}$, where $\Lambda$ is an index set, then the delay of Information $I$, denoted by $delay(I)$, is the maximum difference between the infimum of Reflection Time and the supremum of Occurrence Time of all the atomic information in $A$, i.e.

$$delay(I) = max_{\lambda \in \Lambda}\{inf T_{m\lambda} - sup T_{h\lambda}\}. \tag{4.12}$$

Here, $inf T_{m\lambda}$ and $sup T_{h\lambda}$ denote the infimum of $T_{m\lambda}$ and the supremum of $T_{h\lambda}$ respectively. If there exists a supremum of $T_{h\lambda}$ equaling $+\infty$, then

$$infT_{m\lambda} - supT_{h\lambda} = 0. \tag{4.13}$$

Information delay indicates how fast the Carrier reflects the Ontology. It can be seen from Definition 18 that the value of delay could be positive or negative. Especially, if

$$infT_{m\lambda} < supT_{h\lambda}, \ delay(I) < 0, \tag{4.14}$$

it reflects the Carrier's prediction of future relevant states before the Ontology's State Occurrence Time $T_h$.

**Proposition 6** (Total delay is longer than partial delay) For Information $I = \langle o, T_h, f, c, T_m, g \rangle$ and $I' = \langle o', T'_h, f', c', T'_m, g' \rangle$, if all the atomic information of $I'$ are also the atomic information of $I$, then

$$delay(I') \leq delay(I). \tag{4.15}$$

4.7 The Metric of Information Distribution

The value of Information is often related to its distribution, which is reflected by the coverage of the Carrier. For some information, the wider the coverage is (within a certain extent), the higher the Information value is; and for other information, the narrower the coverage is, the higher the value is. Therefore, in any case Information could be measured by Distribution or Coverage, the mathematical expression for which is:

**Definition 19** (Synonymy Information) Assuming that Information $I_1 = \langle o_1, T_{h1}, f_1, c_1, T_{m1}, g_1 \rangle$ and $I_2 = \langle o_2, T_{h2}, f_2, c_2, T_{m2}, g_2 \rangle$ are sub-information of Information $I = \langle o, T_h, f, c, T_m, g \rangle$, if

$$I^{-1}(g_1(c_1, T_{m1})) = I^{-1}(g_2(c_2, T_{m2})), \tag{4.16}$$

then $I_1$ and $I_2$ are called Synonymy Information in $I$.

**Definition 20** (Information Coverage) Assuming that Information $I_0 = \langle o_0, T_{h0}, f_0, c_0, T_{m0}, g_0 \rangle$ is sub-information of $I = \langle o, T_h, f, c, T_m, g \rangle$, then the coverage of Information $I_0$ in $I$, denoted by $coverage_{I_0}(I)$ is the ratio of the cardinality of the Carrier set of all the sub-information synonymous with $I_0$ in $I$ to the cardinality of Carrier $c$, i.e.

$$coverage_{I_0}(I) = {|\cup_{\lambda \in \Lambda} c_\lambda|}/{|c|}. \tag{4.17}$$

Here, $\{I_\lambda = \langle b_\lambda, T_{h\lambda}, f_\lambda, o_\lambda, T_{m\lambda}, g_\lambda \rangle\}_{\lambda \in \Lambda}$ ($\Lambda$ is an index set) is a set of sub-information synonymous with Information $I_0$.

**Proposition 7** (Partial coverage is wider than total coverage) For Information $I = \langle b, T_h, f, o, T_m, g \rangle$, $I_0 = \langle b_0, T_{h0}, f_0, o_0, T_{m0}, g_0 \rangle$ and $I_1 = \langle b_1, T_{h1}, f_1, o_1, T_{m1}, g_1 \rangle$, if

$$I_0 \subseteq I_1 \subseteq I, \tag{4.18}$$

then

$$coverage_{I_1}(I) \leq coverage_{I_0}(I) \tag{4.19}$$

4.8 The Metric of Information Validity

As the mapping between Ontology State and Carrier State, information should not be labeled true or false, because according to the property of reducibility, there always exists an inverse mapping of information in theory that could restore the actual state of Information Ontology at the Occurrence Time. However, due to the complexity of the mapping process and the existence of subjective factors in the perception process, it is often impossible to acquire the exact inverse mapping of information. Therefore, only by deducing as far as possible can we get close to the actual state of Ontology at the Occurrence Time, i.e.

**Definition 21** (Semantic Mapping and Semantic State) For Information $I = \langle o, T_h, f, c, T_m, g \rangle$, assuming that there is mapping $J$, such that

$$J(g(c, T_m)) = \tilde{f}(\tilde{o}, \widetilde{T_h}), \tag{4.20}$$

where $\tilde{o} \in 2^{O \cup S}$, $\widetilde{T_h} \in 2^T$ and $\tilde{f}(\tilde{o}, \widetilde{T_h})$ is a certain state set of $\tilde{o}$ on $\widetilde{T_h}$, then $J$ is called a Semantic Mapping of $I$, and $\tilde{f}(\tilde{o}, \widetilde{T_h})$ is Semantic State of $I$ with respect to $J$.

The value of Information is to a large degree determined by whether we could acquire the appropriate semantic mapping, such that the semantic state is closest to the actual Ontology State. Generally speaking, when other conditions are all the same, the less the difference between semantic state and the actual state is, the higher the Information value is; and conversely the lower the value is. Therefore, Information could be measured by Validity with respect to semantic mapping, the mathematical expression for which is:

**Definition 22** (Information Validity) For Information $I = \langle o, T_h, f, c, T_m, g \rangle$, assuming that the Semantic States of its State Set $f(o, T_h)$ and Reflection Set $g(c, T_m)$ with respect to semantic mapping $J$, denoted by $\tilde{f}(\tilde{o}, \widetilde{T_h})$, are both elements of distance space $\langle \mathcal{F}, d \rangle$, where $d$ is the distance on $\mathcal{F}$, then the validity of Information $I$ in distance space $\langle \mathcal{F}, d \rangle$ with respect to semantics $J$, denoted by $validity_J(I)$, is the distance between $f(o, T_h)$ and $\tilde{f}(\tilde{o}, \widetilde{T_h})$, i.e.

$$validity_J(I) = d(f, \tilde{f}). \tag{4.21}$$

4.9 The Metric of Information Matchability

The value of Information eventually depends on to what degree it matches users' demand, which is embodied in the overall degree to which the various elements of Information meet users' demand. Generally speaking, the higher the overall degree is, the higher the Information value is; and conversely the lower the value is. Therefore, Information could be measured by Matchability or Suitability, the mathematical expression for which is:

Definition 23 (Information Suitability) For Information $I = \langle o, T_h, f, c, T_m, g \rangle$, assuming that $o, T_h, f, c, T_m, g$ are the elements of the sets $\mathcal{P}_o, \mathcal{P}_{T_h}, \mathcal{P}_f, \mathcal{P}_c, \mathcal{P}_{T_m}, \mathcal{P}_g$ respectively, and $I$ is an element in the distance space $\langle (\mathcal{P}_o, \mathcal{P}_{T_h}, \mathcal{P}_f, \mathcal{P}_c, \mathcal{P}_{T_m}, \mathcal{P}_g), d \rangle$. For the target Sextuple $I_{\lambda_0} = \langle o_{\lambda_0}, T_{h\lambda_0}, f_{\lambda_0}, c_{\lambda_0}, T_{m\lambda_0}, g_{\lambda_0} \rangle$ in Information Sextuples set

$$\{I_\lambda = \langle o_\lambda, T_{h\lambda}, f_\lambda, c_\lambda, T_{m\lambda}, g_\lambda \rangle\}_{b_\lambda \in \mathcal{P}_1, T_{h\lambda} \in \mathcal{P}_2, f_\lambda \in \mathcal{P}_3, o_\lambda \in \mathcal{P}_4, T_{m\lambda} \in \mathcal{P}_5, g_\lambda \in \mathcal{P}_6} (\lambda \in \Lambda, \Lambda \text{ is an index set}), \tag{4.22}$$

the suitability of Information $I$ with respect to target information $I_{\lambda_0}$ is defined as

$$suitability_{I_{\lambda_0}}\big(I(f(b, T_h), o, T_m)\big) = d(I, I_{\lambda_0}). \tag{4.23}$$

The above nine kinds of metrics originate from the Information Sextuple model and have taken into consideration both the inherent properties of and the external needs for Information. Each metric is denoted by quantitative figures, including measure, potential and distance, of specific elements, and meanwhile each element is reflected in several metrics. The entire metric system is relatively comprehensive, and the metrics are independent from each other. Although the definitions of measure, potential and distance are relatively abstract in order to be universally applicable, they could directly correspond to clear and easy mathematical formulas for specific information systems, and therefore could be effectively applied in practical use. For the validity and suitability of information, we have not got the relationship between the measurements of the part and the whole, which shows that neither the validity nor the suitability would change along with the amount of information. This conforms to both common sense and the operation rules of information systems.

5  Conclusions

Only by the objective reflection, i.e. information, can the existence and motion states of everything be reflected. The relationship between different things has to be conducted through Information. Just because of this, information is given the same position as matter and energy, becoming one of the three fundamental elements of the objective world. The importance of information has been fully recognized and a large amount of research achievements in many respects have been reached. American scientist Shannon revealed the basic rules of information transmission and made great contribution to the entry into information society; Zhong put forward the Comprehensive Information Theory and opened the theoretical door into the profound interaction between information and human perception and intelligence [73]. Based on previous studies, this paper tries to explore the essence and connotation of information. Compared with Zhong's theory, this paper converges the essence of information from the full dimension view to the objective origin; and compared with Shannon's theory, it expands the connotation of information from information transmission to all the fields in which information works. It further brings forward the definition of Objective Information, establishes its Sextuple model and nine kinds of metrics and forms the basic theory frame of Objective Information Theory.

The mathematical base for Objective Information Theory is Set Theory, Measure Theory and Topology, which is relatively abstract but could directly and clearly correspond to the statistical and computational methods widely used in daily life. Therefore, the theory is fully applicable to all specific work commonly involved in the informatization process, such as information value analysis, information system operation theory research as well as information system design and evaluation, including big data and stream media that have attracted widespread concern. Like any other theory, its completeness and practicability are to be tested and improved in future applications.

**Acknowledgements**

The authors have had in-depth discussion with Professor Zhong Yixin of Beijing University of Posts and Telecommunications while writing this paper and gained earnest guidance from him. Researcher Zhu Decheng of China Academy of Electronics and Information Technology has also made precious suggestions concerning this paper, and Ms. Li Xinxin and Ms. Fang Fang have offered great help for its composition. Sincere gratitude to all of them!